\documentstyle[12pt]{article}

\def\largelinestretch{\renewcommand{\baselinestretch}{1.265}}

\largelinestretch\small\normalsize
\title{
\vspace*{-10mm}
\hfill
\parbox{4cm}{\large JINR E2-95-238\\
	        hep-ph/9506237}\\
\vspace*{10mm}
        Calculation of Heat-Kernel Coefficients\\
                and Usage of Computer Algebra%
\footnote{
   Talk at the
Fourth International Workshop on Software Engineering and
Artificial Intelligence for High Energy and Nuclear Physics AIHENP'95,
Pisa (Italy), April 3--8, 1995.}}
 \author{
A.A.Bel'kov${}^1$,
A.V.Lanyov${}^1$\thanks{E-mail: {\tt lanyov@cv.jinr.dubna.su}}~,
A.Schaale${}^2$
\\
\\ \normalsize
${}^1$
        Particle Physics Laboratory,\\ \normalsize
 Joint Institute for Nuclear Research,\\ \normalsize
        141980 Dubna, Moscow Region, Russia
\hfill\\ \normalsize
${}^2$
        Deutsches Elektronen-Synchrotron DESY,\\ \normalsize
        Institut f\"ur Hochenergiephysik IfH, \hfill\\ \normalsize
      Platanenallee 6, D-15735 Zeuthen, Germany\hfill\\
}
\date{}
\begin{document}
\largelinestretch\normalsize
\thispagestyle{empty}
\begin{titlepage}
\thispagestyle{empty}
\maketitle
\begin{abstract}
  The calculation of heat-kernel coefficients with
the classical DeWitt algorithm has been discussed.
   We present the explicit form of the coefficients up to
$h_5$ in the general case and up to $h_7^{min}$ for the minimal parts.
   The results are compared with the expressions in other papers.
   A method to optimize the usage of memory
for working with large expressions on universal computer
algebra systems has been proposed.
\end{abstract}
\end{titlepage}

                                \vspace*{20mm}
\section*{Introduction}
   The calculation of the quark determinant modulus
can be reduced to the calculation of the Schwinger proper-time integral
\cite{schwinger}:
\begin{equation}
\log(\det{\bf A}) =
 \int^{\infty }_{1/\Lambda^2}{d\tau\over\tau} {\, \mbox{Tr} \,} K(\tau),
\label{logarithm1}
\end{equation}
where $K(\tau)=e^{-{\bf A}{\tau}}$ is the so-called ``heat kernel''
for operator ${\bf A}={\bf \widehat{D}}^{\dagger}{\bf \widehat{D}}$,
with ${\bf \widehat{D}}\equiv \gamma^\mu {\bf D_\mu}$ being
Dirac differential operator in the presence of the external background fields.
  The coefficients $h_k$ of the expansion of the interaction part
of the heat kernel in
powers of the proper time $\tau$
are known as the heat-kernel coefficients.
   They determine the low-energy expansion of the  effective
one-loop action \cite{ball}.
   Therefore, the calculation of these coefficients
is an alternative method to the direct calculation of Feynman diagrams
\cite{vilkovisky}, thus having a fundamental character.

   This paper is an extension of our previous paper \cite{heat-our}.
   We discuss the implementation of the classical DeWitt algorithm
\cite{dewitt} to calculate the higher-order heat-kernel coefficients by
means of computer algebra.
   After presenting the results obtained with this method we briefly
compare them with those of other groups using different
techniques.
   This method is demonstrated to obtain also the
nonlocal corrections taking into account the finite sizes of mesons,
that leads to the modification of the heat-kernel equation.
   We also discuss technical questions connected with the work with large
expressions in Lisp-based computer algebra systems such as Reduce.
   Our techniques allow us to push the calculations of the
heat-kernel coefficients to higher orders.

\section{General outline of the heat-kernel techniques}

The logarithm of the determinant of a positive definite operator ${\bf A}$ is
defined in proper-time regu\-la\-ri\-zation with the integral
relation Eq.~(\ref{logarithm1}).
   The trace ${\, \mbox{Tr} \,}$ is
to be understood as a space-time integration and a ``normal''
trace over Dirac, color and flavor indices.
   The operator ${\bf A}$ is an elliptic one in usual cases:
$$
{\bf A} \equiv
{\bf \widehat{D}}^{\dagger}{\bf \widehat{D}} =
d_\mu d^\mu + a(x) + \mu^2\,,
$$
where
$d_\mu =\partial_\mu +\Gamma_\mu$,
operator $\Gamma_\mu$ describes vector gauge fields,
$a(x)$ is a local operator without free derivatives.
   The explicit form of these operators depends on the particular model.
   The effective parameter $\mu$ fixes the
regularization in the region  of low momenta.

   The heat kernel $K(\tau)=e^{-{\bf A}\tau}$ satisfies the equation
$${\partial \over \partial \tau} K(\tau) + {\bf A} K(\tau) = 0$$
with the boundary condition
$$K(\tau=0) = 1.$$
   The asymptotic behavior of ${\bf A}$ at short distances
is defined by the ``free'' part
$$
{\bf A}_0 =\partial_\mu \partial^\mu + \mu^2.
$$
Using the ansatz
$$
K = K_0 H
$$
it is convenient to separate from the heat-kernel its ``free'' part $K_0$,
$$
<x| K_0 | y> =
<x| \exp (-(\partial_\mu \partial^\mu+\mu^2)\tau)| y> =
{1\over (4\pi \tau)^2}
           \exp\left\{-\mu^2\tau+{(x-y)^2 \over 4\tau}\right\},
$$
which satisfies equation
$${\partial K_0\over \partial \tau} + {\bf A}_0 K_0 = 0,$$
with the boundary condition
$$ K_{0}(\tau=0) = 1.$$

The ``interaction'' part $H$ of the heat kernel satisfies the equation
$$( {\partial \over \partial\tau} + {1\over\tau} z_\mu d^\mu + d^\mu d_\mu + a)
          H(x,y;\tau) = 0 ,$$
\begin{equation}
H(x,y=x;\tau=0) = 1,
\label{heat-eq}
\end{equation}
with $z_\mu=x_\mu-y_\mu$. The differential operator
$d_\mu$ acts on $x$ only.
Using now an expansion for $H(\tau)$ in powers of $\tau$
$$H(x,y;\tau) = \sum_{k=0}^{\infty} h_{k}(x,y)\cdot \tau^{k}$$
a recursive relation is obtained from Eq.~(\ref{heat-eq}):
\begin{equation}
( n + z_\mu d^\mu  ) h_n(x,y) = - ( a + d_\mu d^\mu  ) h_{n-1}(x,y),
\label{basic-prop}
\end{equation}
with the boundary condition
\begin{equation}
z_\mu d^\mu h_{0}(x,y) = 0.
\label{bound2}
\end{equation}

   After the integration over $\tau$ in Eq.~(\ref{logarithm1}) the
following expression for the quark determinant is obtained
$$
\log |\det i{\bf\widehat{D}}| =
- {1\over 2}\log (\det {\bf \widehat{D}}^{\dagger}{\bf \widehat{D}}) =
- {1\over 2} {\mu^4\over (4\pi)^2}
    \sum_k {\Gamma (k-2,\mu^2/\Lambda^2)\over \mu^{2k}} {\, \mbox{Tr} \,} h_k,
$$
where $\Gamma (\alpha,x)=\int^{\infty }_{x} d t \, e^{-t}t^{\alpha-1}$
is the incomplete gamma function and \mbox{$h_k\equiv h_k(x,y=x)$}.

\section{Recursive algorithm}

In order to construct a recursive algorithm
let us use the basic property of the heat-kernel coefficients providing
the recursive relation
(\ref{basic-prop})
between $h_n(x,y)$ and $h_{n-1}(x,y)$.
  The boundary condition (\ref{bound2}) for $h_0(x,y)$ is determined from
Eq.~(\ref{basic-prop}) with $n=0$ and $h_{-1}(x,y)=0$.
  The task is to find the coincidence limit $h_n\equiv h_n(x,y)\big|_{z=0}$.
  We cannot set $z=0$ in Eq.~(\ref{basic-prop}) directly, since
when acting by differential operator $d_\alpha$ on this equation we get a
nonvanishing contribution:
$d_\alpha(z_\mu d^\mu h_n)\big|_{z=0} =
 g_{\alpha\mu} \cdot d^\mu h_n\big|_{z=0} =
 d_\alpha h_n\big|_{z=0}
$.

One can see that the usage of Eq.~(\ref{basic-prop}) for the calculation
of $h_n$ will produce terms like
$d_\alpha d_\beta \dots h_n(x,y)\big|_{z=0}$.
In order to get the recursive relation for them we apply the product of
$m$ differential operators $d_\alpha d_\beta \dots d_\omega$ on
Eq.~(\ref{basic-prop}) and take limit $z=0$:
\begin{eqnarray}
&&
\underbrace{d_\alpha d_\beta \dots d_\omega}_{m} h_n(x,y)\big|_{z=0} =
\nonumber \\
&&
\hspace{5em} - {1\over n+m}\bigg\{
       d_\alpha d_\beta \dots d_\omega (a+d_\mu d^\mu)h_{n-1}(x,y)
     + P_{\alpha\beta\dots\omega} h_n(x,y)
   \bigg\}\bigg|_{z=0}
\hspace{1em}
\label{basic-rel}
\end{eqnarray}
where $(n+m)>0$ and
$$
P_{\alpha\beta\dots\omega} =
         \underbrace{d_\alpha d_\beta \dots d_\omega}_{m}
           \cdot z_\mu d^\mu\big|_{z=0}
       - m d_\alpha d_\beta \dots d_\omega \, .
$$
For $P_{\alpha\beta\dots\omega}$ we have the recursive relation
\begin{equation}
P_{\alpha\beta\dots\omega} = d_\alpha P_{\beta\dots\omega}
 + R_{\beta\dots\omega; \,\alpha}
\label{p-rel}
\end{equation}
with the boundary condition $P=0$,
where
$R_{\beta\dots\omega; \alpha} = [d_\beta \dots d_\omega, d_\alpha]$.
Performing a commutation of the differential operator
$d_\alpha$ successively through the others
$d_\beta$, ..., $d_\omega$, one can move it to the left side.
  Then the two products with $m$ differential operators cancel each other,
and only terms with $(m-2)$ differential operators are left over.
  Finally we will get the recursive relation
\begin{equation}
R_{\beta\gamma\dots\omega; \,\alpha} =
           \Gamma_{\beta\alpha} \cdot d_\gamma \dots d_\omega
         + d_\beta \cdot R_{\gamma\dots\omega; \,\alpha}
\label{r-rel}
\end{equation}
with the boundary condition $R_{;\alpha}=0$.
   Thus, one has to use the recursive
relation (\ref{basic-rel}) starting from $m=0$ to calculate
$h_n(x,y)\big|_{z=0}$.
   After each iteration it is necessary to commute all
differential operators arising
from $d_\mu d^\mu$ or $P_{\alpha\beta\dots\omega}$ to the right side up
to $h_k(x,y)$ introducing commutators of the type
$$
 S_\mu  = [d_\mu ,a] ,\hspace{0.3cm}
 S_{\mu\nu} = [d_\mu ,S_\nu ] , \hspace{0.3cm}
 S_{\alpha\mu\nu} = [d_\alpha,S_{\mu\nu}],
 \quad \hbox{etc.}
$$
\begin{equation}
 \Gamma_{\mu\nu} = [d_\mu ,d_\nu ] ,\hspace{0.3cm}
 K_{\alpha \mu\nu} = [d_{\alpha },\Gamma_{\mu\nu}] ,\hspace{0.3cm}
 K_{\beta \alpha \mu\nu} = [d_{\beta },K_{\alpha\mu\nu}],
 \quad \hbox{etc.}
\label{commutator}
\end{equation}
$n$ and $m$ change under these iterations  in the following way:
either $(n\to n-1)$,
or     $(m\to m-2)$,
or     $(n\to n-1; m\to m+2)$.
It is easy to show that after $2n$ iterations only
$h_0(x,y)$ remains without differential operators.
  At the end one gets the desired result by setting the limit $z=0$ where
$h_0(x,y)\big|_{z=0}=1$.

   Following the strategy outlined above, the calculation of the
heat-kernel coefficients is straightforward but cumbersome.
   The lengthy calculations can be performed only with
computer support.
   The calculation of the heat-coefficients is a recursive
process which can be done by computer algebra very conveniently
using the recursive relations
 (\ref{basic-rel}-\ref{r-rel})
until substitutions cannot be made any further.

\section{Heat-kernel coefficients}
   The expressions for the heat-kernel coefficients
contain a large number of terms which can be related to each other with the
following transformations:
cyclic properties of trace,
commutator identities
(\ref{commutator})%
,
removing physically redundant total derivatives,
renaming of the dummy indices and
Jacobi identities for symbols $K_{\alpha\mu\nu}$.

   It is necessary to
reduce the expressions to some minimal basis of
linearly in\-de\-pen\-dent terms
   for the final presentation of the results and their
comparison with the results of other papers.
   As soon as we know, it is still not possible for nontrivial
cases to bring all expressions to a canonical form using some
identities.
  The solution of this problem is very important since sometimes it is
easier to perform a calculation in a noncanonical basis
than to bring it to a
canonical form or test its equivalence with the similar
expression in another presentation.
   For example, we had to perform by hand the reduction of the
effective chiral lagrangians of the order $p^6$ in the momentum expansion to
a minimal basis \cite{p6},
after most of the initial ``trivial'' simplifications had been done
automatically with computer algebra.

   So far we do not know an efficient general algorithm which allows to
transform the total heat-kernel coefficients into some unique minimal basis.
   For the reduction of the heat-kernel coefficients to the minimal
basis let us use what is always possible to do --- test
whether two expressions are equivalent or not.
   Here it is sufficient to express all $S_{\mu\dots}$ and
$\Gamma_{\mu\nu\dots}$ according to their definitions in terms
of operators $a$ and $d_\mu$.
   The comparison of such expressions is easier
since the only equivalent transformations here are cyclic
permutations and renaming of dummy indices.
%
%
   If we can test the equivalence of any expressions
with some unknown coefficients,
   then we can expand any expression in the minimal basis
solving the system of linear equations for the coefficients of
this expansion.
   The minimal basis itself can also be constructed by rejecting
the linearly dependent terms and including the linearly independent ones.
   These operations were fully automated by means of the
computer algebra
and they have allowed us to obtain the heat-kernel
coefficients $h_1, \ldots h_5$ in the minimal-basis form:
\begin{eqnarray*}&&
h_0 = 1 ,
\\&&
h_1 = -a,
\\&&
\hspace*{-1.5em}
{\, \mbox{Tr} \,} h_2 = {1\over2} {\, \mbox{Tr} \,} \left\{
   a^{2}
 + {1\over 6} \Gamma_{\mu\nu}^2  \right\},
\\&&
\hspace*{-1.5em}
{\, \mbox{Tr} \,} h_3 = {1\over 6} {\, \mbox{Tr} \,} \bigg\{
- a ^{3}
+{1 \over 2}  S _{\mu }^{2}
-{1 \over 2}   a  \Gamma _{\mu \nu }^{2}
+{1 \over 10}  K _{\nu \nu \mu }^2
-{1 \over 15}  \Gamma _{\mu \nu } \Gamma _{\nu \alpha } \Gamma _{\alpha \mu }
  \bigg\},
\\&&
\hspace*{-1.5em}
{\, \mbox{Tr} \,} h_4 = {1\over 24} {\, \mbox{Tr} \,} \bigg\{
 a ^{4}
+ a ^{2} S _{\mu \mu }
+  \left({4 \over 5} a ^{2} \Gamma _{\mu \nu }^{2}
   +{1 \over 5} \left(a  \Gamma _{\mu \nu }\right)^2\right)
-{2 \over 5}   a  S _{\mu } K _{\nu \nu \mu }
+{1 \over 5}  S _{\mu \mu }^2
\\&&
+{4 \over 15}   a  \Gamma _{\mu \nu } \Gamma _{\nu \rho } \Gamma _{\rho \mu }
+  \left(-{2 \over 5} a  K _{\nu \nu \mu }^2
   +{2 \over 15} S _{\alpha \alpha } \Gamma _{\mu \nu }^{2}
   -{8 \over 15} S _{\beta \gamma } \Gamma _{\gamma \alpha }
 \Gamma _{\alpha \beta }\right)
\\&&
+ \left({17 \over 210} \Gamma _{\mu \nu }^{2} \Gamma _{\alpha \beta }^{2}
   +{2 \over 35} \Gamma _{\mu \nu } \Gamma _{\nu \rho } \Gamma _{\mu
\sigma } \Gamma _{\sigma \rho }
   +{1 \over 105} \Gamma _{\mu \nu } \Gamma _{\nu \rho } \Gamma _{\rho \sigma }
\Gamma _{\sigma \mu }
\right.
\\&&
\left.
   +{1 \over 420} \Gamma _{\mu \nu } \Gamma _{\rho \sigma } \Gamma _{\mu \nu }
\Gamma _{\rho \sigma }\right)
+ {16 \over 105} \left( K _{\mu \alpha \alpha \nu } \Gamma _{\nu \rho } \Gamma
_{\rho \mu }
                       +K _{\alpha \alpha \mu } K _{\beta \beta \nu } \Gamma
_{\mu \nu }
                 \right)
+{1 \over 35}  K _{\mu \alpha \alpha \nu }^2
 \bigg\},
\\&&
\hspace*{-1.5em}
{\, \mbox{Tr} \,} h_5 = {1\over 240} {\, \mbox{Tr} \,} \bigg\{      
-a ^{5}
+ \left(-2 a ^{3} S _{\mu \mu }
   -a ^{2} S _{\mu }^{2}\right)
+  \left(-a ^{3} \Gamma _{\mu \nu }^{2}
   -{2 \over 3} a ^{2} \Gamma _{\mu \nu } a  \Gamma _{\mu \nu }\right)
\\&&
+  \left({2 \over 3} a ^{2} S _{\mu } K _{\nu \nu \mu }
   -{2 \over 3} S _{\mu } S _{\nu } a  \Gamma _{\mu \nu }\right)
+ \left(-a  S _{\mu \mu } S _{\nu \nu }
   -{2 \over 3} S _{\mu \mu } S _{\nu }^{2}\right)
\\&&
+  \left(-{2 \over 7} a ^{2} \Gamma _{\mu \nu } \Gamma _{\nu \alpha } \Gamma
_{\alpha \mu }
   -{8 \over 21} a  \Gamma _{\mu \nu } a  \Gamma _{\nu \alpha } \Gamma _{\alpha
\mu }\right)
\\&&
+  \left({4 \over 7} a ^{2} K _{\mu \mu \nu } K _{\alpha \alpha \nu }
   +{3 \over 7} a  K _{\mu \mu \nu } a  K _{\alpha \alpha \nu }
   -{8 \over 7} a  S _{\mu \mu } \Gamma _{\nu \alpha }^{2}
   +{4 \over 7} a  S _{\mu } K _{\nu \nu \alpha } \Gamma _{\mu \alpha }
\right. \\&& \left. \quad
   +{8 \over 7} a  S _{\mu } \Gamma _{\mu \nu } K _{\alpha \alpha \nu }
   -{4 \over 21} a  \Gamma _{\mu \nu } S _{\alpha \alpha } \Gamma _{\mu \nu }
   -{11 \over 21} S _{\mu }^{2} \Gamma _{\nu \alpha }^{2}
   +{20 \over 21} S _{\mu } K _{\nu \nu \alpha } a  \Gamma _{\mu \alpha }
\right. \\&& \left. \quad
   +{2 \over 21} S _{\mu } S _{\nu } \Gamma _{\mu \alpha } \Gamma _{\alpha \nu
}
   -{10 \over 21} S _{\mu } S _{\nu } \Gamma _{\nu \alpha } \Gamma _{\alpha \mu
}
   +{2 \over 7} S _{\mu } \Gamma _{\mu \nu } S _{\alpha } \Gamma _{\alpha \nu }
   +{1 \over 42} S _{\mu } \Gamma _{\nu \alpha } S _{\mu } \Gamma _{\nu \alpha
}\right)
\\&&
+  \left({8 \over 21} S _{\mu \mu } S _{\nu } K _{\alpha \alpha \nu }
   -{4 \over 21} S _{\mu } S _{\nu } K _{\mu \alpha \alpha \nu }\right)
\\&&
+{1 \over 14}  S _{\mu \nu \nu } S _{\mu \alpha \alpha }
+  \left(-{17 \over 84} a  \Gamma _{\mu \nu }^{2} \Gamma _{\alpha \beta }^{2}
   -{1 \over 21} a  \Gamma _{\mu \nu } \Gamma _{\nu \alpha } \Gamma _{\mu \beta
} \Gamma _{\beta \alpha }
   -{1 \over 21} a  \Gamma _{\mu \nu } \Gamma _{\nu \alpha } \Gamma _{\alpha
\beta } \Gamma _{\beta \mu }
\right. \\&& \left. \quad
   -{5 \over 84} a  \Gamma _{\mu \nu } \Gamma _{\alpha \beta }^{2} \Gamma _{\mu
\nu }
   -{13 \over 84} a  \Gamma _{\mu \nu } \Gamma _{\alpha \beta } \Gamma _{\mu
\nu } \Gamma _{\alpha \beta }
   -{5 \over 21} a  \Gamma _{\mu \nu } \Gamma _{\alpha \beta } \Gamma _{\beta
\nu } \Gamma _{\alpha \mu }\right)
\\&&
+  \left(-{2 \over 21} a  K _{\mu \mu \nu } \Gamma _{\nu \alpha } K _{\beta
\beta \alpha }
   -{2 \over 7} a  \Gamma _{\mu \nu } K _{\mu \alpha \beta } K _{\nu \alpha
\beta }
   -{4 \over 21} a  \Gamma _{\mu \nu } K _{\nu \alpha \alpha \beta } \Gamma
_{\beta \mu }
   -{2 \over 21} a  \Gamma _{\mu \nu } K _{\alpha \alpha \mu } K _{\beta \beta
\nu }
\right. \\&& \left. \quad
   -{4 \over 21} S _{\mu \mu } \Gamma _{\nu \alpha } \Gamma _{\alpha \beta }
\Gamma _{\beta \nu }
   -{4 \over 21} S _{\mu \nu } \Gamma _{\mu \alpha } \Gamma _{\beta \nu }
\Gamma _{\alpha \beta }
   +{2 \over 7} S _{\mu } K _{\nu \nu \mu } \Gamma _{\alpha \beta }^{2}
   -{2 \over 7} S _{\mu } K _{\nu \nu \alpha } \Gamma _{\mu \beta } \Gamma
_{\beta \alpha }
\right. \\&& \left. \quad
   +{2 \over 21} S _{\mu } \Gamma _{\mu \nu } K _{\alpha \alpha \beta } \Gamma
_{\beta \nu }\right)
+  \left(-{1 \over 7} a  K _{\mu \nu \nu \alpha } K _{\mu \beta \beta \alpha }
   +{2 \over 21} a  K _{\mu \nu \nu \alpha } K _{\alpha \beta \beta \mu }
   -{3 \over 28} S _{\mu \mu \nu \nu } \Gamma _{\alpha \beta }^{2}
\right. \\&& \left. \quad
   -{1 \over 42} S _{\mu \mu } K _{\nu \nu \alpha } K _{\beta \beta \alpha }
   -{2 \over 7} S _{\mu \mu } K _{\nu \alpha \alpha \beta } \Gamma _{\nu \beta
}
   +{1 \over 7} S _{\mu } K _{\nu \nu \alpha \alpha \beta } \Gamma _{\mu \beta
}\right)
\\&&
+ \left(-{47 \over 126} \Gamma _{\mu \nu }^{2} \Gamma _{\alpha \beta } \Gamma
_{\beta \gamma } \Gamma _{\gamma \alpha }
   -{11 \over 189} \Gamma _{\mu \nu } \Gamma _{\nu \alpha } \Gamma _{\mu \beta
} \Gamma _{\alpha \gamma } \Gamma _{\gamma \beta
}
   +{1 \over 63} \Gamma _{\mu \nu } \Gamma _{\nu \alpha } \Gamma _{\mu \beta }
\Gamma _{\beta \gamma } \Gamma _{\gamma \alpha }
\right. \\&& \left. \quad
   +{37 \over 945} \Gamma _{\mu \nu } \Gamma _{\nu \alpha } \Gamma _{\alpha
\beta } \Gamma _{\beta \gamma } \Gamma _{\gamma \mu }
   +{1 \over 126} \Gamma _{\mu \nu } \Gamma _{\nu \alpha } \Gamma _{\beta
\gamma } \Gamma _{\alpha \mu } \Gamma _{\beta \gamma }
   +{1 \over 945} \Gamma _{\mu \nu } \Gamma _{\alpha \beta } \Gamma _{\gamma
\mu } \Gamma _{\nu \alpha } \Gamma _{\beta \gamma }
\right)
\\&&
+ \left(-{8 \over 189} K _{\mu \mu \nu } K _{\nu \alpha \beta } \Gamma _{\beta
\gamma } \Gamma _{\gamma \alpha }
   -{10 \over 189} K _{\mu \mu \nu } K _{\alpha \alpha \nu } \Gamma _{\beta
\gamma }^{2}
   +{2 \over 21} K _{\mu \mu \nu } K _{\alpha \alpha \beta } \Gamma _{\nu
\gamma } \Gamma _{\gamma \beta }
\right. \\&& \left. \quad
   +{4 \over 63} K _{\mu \mu \nu } \Gamma _{\nu \alpha } K _{\beta \beta \gamma
} \Gamma _{\gamma \alpha }
   +{5 \over 378} K _{\mu \mu \nu } \Gamma _{\alpha \beta } K _{\gamma \gamma
\nu } \Gamma _{\alpha \beta }
   -{61 \over 189} K _{\mu \nu \nu \alpha } \Gamma _{\mu \alpha } \Gamma
_{\beta \gamma }^{2}
\right. \\&& \left. \quad
   +{22 \over 189} K _{\mu \nu \nu \alpha } \Gamma _{\mu \beta } \Gamma
_{\gamma \alpha } \Gamma _{\beta \gamma }
   -{16 \over 189} K _{\mu \nu \nu \alpha } \Gamma _{\alpha \beta } \Gamma
_{\beta \gamma } \Gamma _{\gamma \mu }
   -{10 \over 189} K _{\mu \nu \nu \alpha } \Gamma _{\alpha \beta } \Gamma
_{\gamma \mu } \Gamma _{\beta \gamma }
\right. \\&& \left. \quad
   -{2 \over 189} K _{\mu \nu \nu \alpha } \Gamma _{\beta \gamma } \Gamma _{\mu
\alpha } \Gamma _{\beta \gamma }
   -{4 \over 63} K _{\mu \nu \alpha }^{2} \Gamma _{\beta \gamma }^{2}
   -{4 \over 189} K _{\mu \nu \alpha } K _{\mu \alpha \beta } \Gamma _{\nu
\gamma } \Gamma _{\gamma \beta }
\right. \\&& \left. \quad
   -{4 \over 189} K _{\mu \nu \alpha } K _{\mu \alpha \beta } \Gamma _{\beta
\gamma } \Gamma _{\gamma \nu }
   -{2 \over 63} K _{\mu \nu \alpha } K _{\beta \nu \alpha } \Gamma _{\mu
\gamma } \Gamma _{\gamma \beta }
   +{2 \over 63} K _{\mu \nu \alpha } \Gamma _{\nu \alpha } K _{\mu \beta
\gamma } \Gamma _{\beta \gamma }
\right. \\&& \left. \quad
   +{2 \over 189} K _{\mu \nu \alpha } \Gamma _{\alpha \beta } K _{\mu \nu
\gamma } \Gamma _{\gamma \beta }\right)
+ \left(-{1 \over 42} K _{\mu \mu \nu \nu \alpha } K _{\alpha \beta \gamma }
\Gamma _{\beta \gamma }
   +{1 \over 63} K _{\mu \nu \nu \alpha } K _{\mu \beta \beta \gamma } \Gamma
_{\alpha \gamma }
\right. \\&& \left. \quad
   +{4 \over 63} K _{\mu \nu \nu \alpha } K _{\alpha \beta \beta \gamma }
\Gamma _{\mu \gamma }
   -{ 5\over 63} K _{\mu \nu \nu \alpha } K _{\beta \beta \mu } K _{\gamma
\gamma \alpha }\right)
+{1 \over 126}  K _{\mu \mu \nu \nu \alpha } K _{\beta \beta \gamma \gamma
\alpha }
 \bigg\} + \mbox{h.c}.
\end{eqnarray*}
Here Hermitian conjugation h.c.~is defined by
$$a^\dagger=a,
  \quad (S_{\mu\ldots})^\dagger=S_{\mu\ldots},
$$
$$(\Gamma_{\mu\nu})^\dagger = -\Gamma_{\mu\nu}, \quad
  (K_{\alpha\mu\nu\ldots})^\dagger = -K_{\alpha\mu\nu\ldots} .
$$
Our expressions for the heat-kernel coefficients $h_{4,5}$
proved to be equivalent with the results of
Refs.~\cite{van-de-ven,avramidi}.
   On the other hand, we have found deviations
from Ref.~\cite{ball} in the part containing terms with 6 and 8 indices:
\begin{eqnarray*}
&&
   (24{\, \mbox{Tr} \,} h_4)_{\mbox{this Ref.}}
 - (24{\, \mbox{Tr} \,} h_4)_{\mbox{Ref.~\cite{ball}}}
\\&&
=
\left(-{8 \over 15} a  d _{\mu }^{2} d_{\nu } d _{\alpha }^{2} d _{\nu }
   +{8 \over 15} a  d _{\mu }^{2} d _{\nu } d _{\alpha } d _{\nu } d _{\alpha }
   -{8 \over 15} a  d _{\mu } d _{\nu }^{2} d _{\mu } d _{\alpha }^{2}
   +{8 \over 15} a  d _{\mu } d _{\nu } d _{\mu } d _{\nu } d _{\alpha }^{2}
\right. \\&& \left. \quad
   +{16 \over 15} a  d _{\mu } d _{\nu } d _{\alpha }^{2} d _{\nu } d _{\mu }
   -{16 \over 15} a  d _{\mu } d _{\nu } d _{\alpha } d _{\nu } d _{\alpha } d
   _{\mu }\right)
\\&&
-{16 \over 105} d _{\mu }^{2} d _{\nu }^{2} d _{\alpha }^{2} d _{\beta }^{2}
-{16 \over 105} d _{\mu }^{2} d _{\nu }^{2} d _{\alpha } d _{\beta }^{2}
d_{\alpha }
+{64 \over 105} d _{\mu }^{2} d _{\nu }^{2} d _{\alpha } d _{\beta } d _{\alpha
 } d _{\beta }
+{32 \over 105} d _{\mu }^{2} d _{\nu } d _{\alpha } d _{\nu } d _{\beta }^{2}
d_{\alpha }
\\&&
-{64 \over 105} d _{\mu }^{2} d _{\nu } d _{\alpha } d _{\nu } d _{\beta }
d_{\alpha } d _{\beta }
+{16 \over 105} d _{\mu }^{2} d _{\nu } d _{\alpha } d _{\beta }^{2} d _{\nu }
d_{\alpha }
+{32 \over 105} d _{\mu }^{2} d _{\nu } d _{\alpha } d _{\beta } d _{\nu }
d_{\alpha } d _{\beta }
\\&&
-{16 \over 35} d _{\mu }^{2} d _{\nu } d _{\alpha } d _{\beta } d _{\nu }
d_{\beta } d _{\alpha }
-{16 \over 35} d _{\mu }^{2} d _{\nu } d _{\alpha } d _{\beta } d _{\alpha } d
_{\nu } d _{\beta }
+{16 \over 105} d _{\mu }^{2} d _{\nu } d _{\alpha } d _{\beta } d _{\alpha }
d_{\beta } d _{\nu }
\\&&
-{32 \over 105} d _{\mu } d _{\nu } d _{\mu } d _{\nu } d _{\alpha } d _{\beta
}d _{\alpha } d _{\beta }
+{64 \over 105} d _{\mu } d _{\nu } d _{\mu } d _{\alpha } d _{\nu } d _{\beta
}d _{\alpha } d _{\beta }
-{32 \over 105} d _{\mu } d _{\nu } d _{\mu } d _{\alpha } d _{\beta } d _{\nu
}d _{\alpha } d _{\beta }
\\&&
+{32 \over 105} d _{\mu } d _{\nu } d _{\mu } d _{\alpha } d _{\beta } d _{\nu
}d _{\beta } d _{\alpha }
\end{eqnarray*}
These deviations arise from the terms of the expression from Ref.~\cite{ball}
$$(+{2\over  5}S_{\alpha\alpha } \Gamma_{\mu\nu }^{2}
   +{4\over 105}
    \{\Gamma_{\alpha\beta} \mid K_{\rho\rho\mu} \mid K_{\mu\alpha\beta}\}),
$$
where $\{A \mid B \mid C\} \equiv ABC+CBA$.
The corresponding terms of our expression are
$$(+{2\over 15}S_{\alpha\alpha } \Gamma_{\mu\nu }^{2}
   +{16\over 105}K_{\mu\alpha\alpha\nu} \Gamma_{\nu\rho} \Gamma_{\rho\mu}).
$$

  We have performed the following checks of our results.
  First of all, we have used the identities
$\partial({\, \mbox{Tr} \,} h_n)/\partial a=-h_{n-1}$ \cite{ball}.
   Secondly,
substituting the explicit form of operators $a$ and
$\Gamma_\mu$ in the
Nambu--Jona-Lasinio (NJL) model \cite{njl}
and evaluating trace over Dirac
gamma-matrices we derived the effective chiral $p^4$- and $p^6$-lagrangians
describing the low-energy processes with mesons and photons \cite{p6}.
   In particular, we reproduce the structure coefficients
$L_i$ of the $p^4$-lagrangians obtained in Ref.~\cite{espriu}.
   We also reproduce the effective Euler-Heisenberg lagrangian
\cite{euler-heisenberg} describing the photon-photon scattering
for particles of both spin 0 and spin $1/2$ \cite{schwinger-lectures}.

  In the case $\Gamma_\mu=0$ we can also present the next orders
of the ``minimal'' parts of the heat-kernel coefficients:
\begin{eqnarray*}&&
\hspace*{-1.5em}
{\, \mbox{Tr} \,} h_5^{min} =
{1\over 120} {\, \mbox{Tr} \,} \Bigg\{
- a ^{5}
+ 3 a ^{2} S _{\mu }^{2}
+ 2 a  S _{\mu } a  S _{\mu }
- a  S _{\mu \nu }^{2}
- {5 \over 3} S _{\mu } S _{\nu } S _{\mu \nu }
+ {1 \over 14}  S _{\mu \nu \alpha }^2
\Bigg\},
\\&&
\hspace*{-1.5em}
{\, \mbox{Tr} \,} h_6^{min} = {1\over 720} {\, \mbox{Tr} \,} \Bigg\{
a ^{6}
+ \left(-4 a ^{3} S _{\mu }^{2}
   -6 a ^{2} S _{\mu } a  S _{\mu }  \right)
+ \left({12 \over 7} a ^{2} S _{\mu \nu }^{2}
   +{9 \over 7} a  S _{\mu \nu } a  S _{\mu \nu }
\right. \\&& \left. \quad
   +{26 \over 7} a  S _{\mu \nu } S _{\mu } S _{\nu }
   +{18 \over 7} a  S _{\mu } S _{\mu \nu } S _{\nu }
   +{26 \over 7} a  S _{\mu } S _{\nu } S _{\mu \nu }
   +{9 \over 7} S _{\mu }^{2} S _{\nu }^{2}
   +{17 \over 14} S _{\mu } S _{\nu } S _{\mu } S _{\nu }  \right)
\\&&
+ \left(-{3 \over 7} a  S _{\mu \nu \alpha }^{2}
   -{11 \over 21} S _{\mu \nu } S _{\nu \alpha } S _{\mu \alpha }
   -S _{\mu } S _{\mu \nu \alpha } S _{\nu \alpha }
   -S _{\mu } S _{\nu \alpha } S _{\mu \nu \alpha }    \right)
+{1 \over 42}  S _{\mu \nu \alpha \beta }^{2}
\Bigg\},
\\&&
\hspace*{-1.5em}
{\, \mbox{Tr} \,} h_7^{min} = {1\over 5040} {\, \mbox{Tr} \,} \Bigg\{
- a ^{7}
+ \left(5 a ^{4} S _{\mu }^{2}
   +8 a ^{3} S _{\mu } a  S _{\mu }
   +{9 \over 2} a ^{2} S _{\mu } a ^{2} S _{\mu }  \right)
\\&&
+ \left(-{5 \over 2} a ^{3} S _{\mu \nu }^{2}
   -{9 \over 2} a ^{2} S _{\mu \nu } a  S _{\mu \nu }
   -6 a ^{2} S _{\mu \nu } S _{\mu } S _{\nu }
   -{7 \over 2} a ^{2} S _{\mu } S _{\mu \nu } S _{\nu }
   -6 a ^{2} S _{\mu } S _{\nu } S _{\mu \nu }
\right. \\&& \left. \quad
   -{7 \over 2} a  S _{\mu }^{2} S _{\nu }^{2}
   -{11 \over 2} a  S _{\mu } a  S _{\mu \nu } S _{\nu }
   -{11 \over 2} a  S _{\mu } a  S _{\nu } S _{\mu \nu }
   -{11 \over 2} a  S _{\mu } S _{\nu }^{2} S _{\mu }
   -{17 \over 2} a  S _{\mu } S _{\nu } a  S _{\mu \nu }
\right. \\&& \left. \quad
   -{17 \over 2} a  S _{\mu } S _{\nu } S _{\mu } S _{\nu }  \right)
+ \left({5 \over 6} a ^{2} S _{\mu \nu \alpha }^{2}
   +{2 \over 3} a  S _{\mu \nu \alpha } a  S _{\mu \nu \alpha }
   +{17 \over 6} a  S _{\mu \nu \alpha } S _{\mu } S _{\nu \alpha }
\right. \\&& \left. \quad
   +{5 \over 2} a  S _{\mu \nu \alpha } S _{\nu \alpha } S _{\mu }
   +{5 \over 3} a  S _{\mu \nu } S _{\mu \nu \alpha } S _{\alpha }
   +{11 \over 3} a  S _{\mu \nu } S _{\nu \alpha } S _{\mu \alpha }
   +{17 \over 6} a  S _{\mu \nu } S _{\alpha } S _{\mu \nu \alpha }
\right. \\&& \left. \quad
   +{5 \over 3} a  S _{\mu } S _{\mu \nu \alpha } S _{\nu \alpha }
   +{5 \over 2} a  S _{\mu } S _{\nu \alpha } S _{\mu \nu \alpha }
   +{5 \over 3} S _{\mu }^{2} S _{\nu \alpha }^{2}
   +{35 \over 18} S _{\mu } S _{\mu \nu } S _{\alpha } S _{\nu \alpha }
\right. \\&& \left. \quad
   +{11 \over 6} S _{\mu } S _{\nu \alpha } S _{\mu } S _{\nu \alpha }
   +{35 \over 18} S _{\mu } S _{\nu \alpha } S _{\alpha } S _{\mu \nu }
   +{97 \over 18} S _{\mu } S _{\nu } S _{\mu \alpha } S _{\nu \alpha }
   +{43 \over 18} S _{\mu } S _{\nu } S _{\nu \alpha } S _{\mu \alpha }
\right. \\&& \left. \quad
   +{35 \over 9} S _{\mu } S _{\nu } S _{\alpha } S _{\mu \nu \alpha } \right)
+ \left(-{1 \over 6} a  S _{\mu \nu \alpha \beta }^{2}
   -{16 \over 15} S _{\mu \nu } S _{\mu \alpha \beta } S _{\nu \alpha \beta }
   -{7 \over 10} S _{\mu \nu } S _{\alpha \beta } S _{\mu \nu \alpha \beta }
\right. \\&& \left. \quad
   -{1 \over 2} S _{\mu } S _{\mu \nu \alpha \beta } S _{\nu \alpha \beta }
   -{1 \over 2} S _{\mu } S _{\nu \alpha \beta } S _{\mu \nu \alpha \beta }
 \right)
+{1 \over 132}  S _{\mu \nu \alpha \beta \gamma }^{2}
\Bigg\}.
\end{eqnarray*}

The expressions for the heat-kernel coefficients up to
${\, \mbox{Tr} \,} h_6^{min}$ have been also presented in Ref.~\cite{carson},
and up to ${\, \mbox{Tr} \,} h_8^{min}$ in Ref.~\cite{fliegner}.
   The minimal coefficients $h_n^{min}$ can be
easily transformed into a unique minimal basis \cite{fliegner}:
it is necessary to move all identical indices in each
symbol, e.g.\ $S_{\mu\nu\mu\dots}$, to the rest symbols in every term
because any addition of some total derivative to the lagrangian
does not alter its physical content.
   This way we have checked our expressions for the
minimal coefficients to coincide with other papers.

\section{Nonlocal corrections for the effective lagrangian}
   The usage of the recursive relations allows us to extend the local
NJL model \cite{njl}
by taking into account the finite sizes of mesons
\cite{bilocal} in a simple way.
   In this case the modulus squared of Dirac operator ${\bf \widehat{D}}$
receives some additional contributions proportional to
a small parameter $\alpha/\Lambda^2$ characterizing the size of
the nonlocal corrections:
\begin{eqnarray*}
{\bf \widehat{D}}^\dagger{\bf \widehat{D}}
&=& \beta \partial^2 + \mu^2
  + 2\Gamma_{\mu}\partial^{\mu} + \Gamma_{\mu}^2 + a
\nonumber \\
  &&+ \frac{\alpha}{\Lambda^2}\Big[ b
  +Q_{\alpha}\partial^{\alpha}
  + (a+c)\partial^2 + 2\big(\Gamma_{\mu}\partial^2
                           +\partial_{\alpha}\Gamma_{\mu}\partial^{\alpha}
                      \big)\partial^{\mu} \Big]
  + O\bigg(\frac{\alpha^2}{\Lambda^4} \bigg),
\end{eqnarray*}
where $\beta = 1+2\alpha\mu^2 / \Lambda^2$
and $a$, $b$, $c$ and $Q_\mu$ are local operators
(their explicit form can be found in Ref.~\cite{bilocal}).

The modified recursive equation
for the heat-kernel coefficients $h_n$
(for simplicity we study only the case without vector fields)
reads
\begin{eqnarray*}&&
\hspace*{-1.5em}
  \frac{\alpha}{4\Lambda^2} z^2 c h_{n+1}(x,y)
\nonumber \\&&
\hspace*{-1.5em}
  +\bigg\{ n + z_\mu \partial^\mu + \frac{\alpha}{\Lambda^2}
\bigg[
    2(a+c)
    \big(1+{1\over2}z_\mu \partial^\mu\big)
    -2 \mu^2 z_\mu \partial^\mu
    +{1\over2} z_\mu\big(\partial^\mu a + 2Q^\mu\big)
    \bigg] \bigg\} h_n(x,y)
\nonumber \\&&
\hspace*{-1.5em}
    + \bigg\{a + \partial^2 + \frac{\alpha}{\Lambda^2}\bigg[
      b
    + \big(\partial_{\mu}
      +2Q^{\mu} \big) \partial^{\mu}
    + (a+c) \partial^2 \bigg)\bigg]\bigg\} h_{n-1}(x,y) =0\, .
\label{heat-eq-bil}
\end{eqnarray*}
Analogously to the algorithm described above, one can obtain the
recursive relations for
$d_\alpha d_\beta \dots h_n(x,y)\big|_{z=0}$, where each
iteration gives either
   $(n\to n-1)$,
or $(m\to m-1)$,
or $(m\to m-2)$,
or $(n\to n-1; m\to m+1)$,
or $(n\to n-1; m\to m+2)$.
It is easy to see that functional $F=3n+m$ reduces at least by~1
after each iteration.
  Therefore, the final result is obtained after $3n$ iterations \cite{bilocal}.

\section{Implementation of calculation in computer algebra system Reduce}

We have used the Computer Algebra System (CAS)
Reduce \cite{reduce} extended by our package for the calculations in
chiral meson theories \cite{reduce-package}.
   Reduce suits well for our purposes since it is a universal and open
system which can be easily extended by the users.
   It is widely used for our calculations of amplitudes of various
meson processes and for the derivation and transformation of lagrangians
from bosonization of the NJL model.
   It allows us to complete the general mathematical and algorithmic
   base of Reduce by the specific data types and operations for this field.
   In general it has provided a convenient computational environment
for studying chiral meson theories.
   A more detailed description of our package can be found
in Ref.~\cite{reduce-package}.

   For the calculations described in this paper the following package features
are the most essential:
\begin{itemize}
\item
   transformation by cyclic permutation of
   noncommutative operator products under the trace operation,
   e.g.~${\, \mbox{Tr} \,}(ABC)\equiv{\, \mbox{Tr} \,}(BCA)$;
\item
   transformation by redefinition of dummy indices,
   e.g.~$S_{\mu\nu}S_\mu S_\nu \equiv S_{\nu\mu}S_\nu S_\mu$;
\item
   \LaTeX{} output of large expressions
   (on the basis of RLFI package from Reduce library \cite{rlfi}).
\end{itemize}
   For the work with the heat-kernel coefficients
the following operations with indices have been additionally introduced:
\begin{itemize}
\item
ordering indices in a symbol%
:
$$S_{\nu\mu}\to S_{\mu\nu} + a\Gamma_{\mu\nu} - \Gamma_{\mu\nu}a;$$
\item
moving indices from the symbols with too many of them%
:
$${\, \mbox{Tr} \,}(S_{\alpha\mu\nu}S_\alpha S_\mu S_\nu)\to
-{\, \mbox{Tr} \,}(S_{\mu\nu}[d_\alpha, S_\alpha S_\mu S_\nu]);$$
\item
moving apart identical indices to different symbols%
:
$${\, \mbox{Tr} \,}(S_{\mu\mu}S_{\nu\nu})\to{\, \mbox{Tr}
\,}(S_{\mu\nu}S_{\mu\nu})
                                                +\mbox{commutator terms}.
$$
\end{itemize}
   As it was already noted, these procedures are not sufficient
to transform expressions to some unique form.
   Nevertheless, they perform the overwhelming part of the work to
reduce the original expressions.

   Universal Lisp-based CAS such as Reduce cannot work with large
expressions which do not fit into available computer memory.
  At
the same time the intermediate expressions can grow very much during
the large-scale recursive calculations.
  This problem is effectively tackled by the CAS Form \cite{form} which works
with the terms of large expressions ``locally'' keeping only
a relatively small part of the terms in the memory and storing the
others on the disk.
%
%
   Another alternative is writing a specialized program to
calculate the heat-kernel coefficients in the language C \cite{kornyak}.

We have
implemented the following simple method
to resolve the problem of uncontrolled growth of expressions
causing the memory overflow in Reduce:
   During the work of algorithm the expression is regularly (after
each iteration) looked through and a limit is imposed on the
number of terms to be substituted at the next iteration.
  This can be done by a simple change of all differential operators
$d_\mu$, except the first $n_{max}$ ones, to some new operator
$\tilde{d_\mu}$ for which no substitutions are set.
   The parameter $n_{max}$ is chosen to provide an optimal
loading of memory without its overflow.
   This method has turned out to be effective.
It allowed us to calculate the next orders of the heat-kernel coefficients.

\section*{Conclusion}
   The usage of the classical DeWitt algorithm allows to
generate effective lagrangians for a wide range of
problems in a simple way.
   In our previous papers \cite{p6}
we obtained an effective chiral lagrangian from bosonization
of the NJL model at $O(p^6)$ order using this method.
   It also allows to study the nonlocal corrections for the $p^4$ chiral
lagrangian \cite{bilocal}.
   The usage of the CAS Reduce extended by a specialized package
allowed us to obtain the higher-order heat-kernel coefficients,
reduce these expressions to minimal basis
and compare the results with some other papers.

\section*{Acknowledgements}
   The authors are grateful to S.Scherer for careful reading
this paper and useful comments.
   The authors would like to thank D.Fliegner, V.P.Gerdt, V.V.Kornyak
   and U.M\"uller
for discussions and B.R.Holstein for correspondence.
   A.A.Bel'kov and A.V.Lanyov are grateful for the hospitality
and support while working at these problems at DESY-Zeuthen.
   This work was supported by the Russian Foundation for
Fundamental Research under grant No.~94-02-03973.
   The participation at the
Fourth International Workshop on Software Engineering and
Artificial Intelligence for High Energy and Nuclear Physics AIHENP'95,
Pisa (Italy), April 3--8, 1995,
was supported by the INFN.

\end{document}